\documentclass[runningheads,natbib]{svjour2}
\bibpunct{[}{]}{;}{n}{}{,} % to get "[numbered]" references from natbib
\smartqed  % flush right qed marks, e.g. at end of proof
\usepackage{graphicx}
\usepackage{mathptmx}      % use Times fonts if available on your TeX system
\journalname{ISSI Scientific Report}
\begin{document}
\begin{sloppypar}

\title{Hard X-ray and $\gamma$-ray Detectors}
%\titlerunning{High-energy Detectors} 
\author{David M. Smith}
\institute{D. M. Smith \at
              Physics Department and Santa Cruz Institute for Particle Physics \\
              University of California, Santa Cruz, USA\\
}

\date{Received: date}
% The correct date will be entered by the editor

\maketitle

\begin{center}
{\bf ABSTRACT}
\end{center}

The detection of photons above 10 keV through MeV and GeV energies is
challenging due to the penetrating nature of the radiation, which can
require large detector volumes, resulting in correspondingly high background.
In this energy range, most detectors in space are either scintillators or
solid-state detectors.  The choice of detector technology depends on
the energy range of interest, expected levels of signal and
background, required energy and spatial resolution, particle
environment on orbit, and other factors.  This section covers the
materials and configurations commonly used from 10~keV to
$>$1~GeV.

\section{Introduction}
\label{intro}

Most high energy detectors in space are based on scintillators or
solid-state detectors.  Scintillators are the older technology and are
generally used where large detector volumes and lower cost are
paramount.  Solid-state diode detectors, e.g.  germanium, silicon, and
cadmium telluride (CdTe) or cadmium zinc telluride (Cd$_{1-x}$Zn$_x$Te
or CZT), generally have better energy resolution than scintillators
and a lower energy threshold, and can be more easily pixellated for
fine spatial resolution if that is required.  They are more expensive
than scintillators and more difficult to produce, package and read out
in large volumes.

For all these materials, photons are detected when their energy is
transferred to electrons via photoelectric absorption, Compton
scattering, or pair production (which also produces positrons, of
course).  The high energy particles come to a stop in the detector
volume, producing ionization that is detected by varying methods.

I will review the most commonly used solid-state and scintillator
materials and detector configurations.  Considerable work has also
been done worldwide on gas and liquid detectors and on newer or less
common semiconductors and scintillators.  For a much more detailed
discussion of a wider range of detectors, as well as an excellent
treatment of general considerations in photon counting and
electronics, see the excellent textbook by \citet{Kn00}.

\section{Configurations and energy regimes}
\label{config}

\subsection{Thin and monolithic detectors}
\label{mono}

The optimum configuration and material for a detector depend most
strongly on the energy range of the photons to be observed.  Figure~1
shows the cross-sections for photoelectric absorption, Compton
scattering, and pair production for photons by elements commonly used
in detectors: silicon and germanium (in solid-state detectors) and
iodine and bismuth (in common scintillators).

Simple efficiency calculations based on cross-sections can assist with
instrument design, particularly when photoelectric interactions are
dominant, but Monte Carlo simulation is the most powerful and flexible
tool.  It can be used to model the response to source and background
radiation and to incident particles other than photons as well.  The
packages most commonly used are GEANT3 and GEANT4 (GEometry
ANd Tracking), originally
developed at the European Organization for Nuclear Research
(CERN) for accelerator applications \citep{Ce93,Ag03}.

Since the cross-section for photoelectric absorption is large at low
energies, low-energy detectors can be quite thin.  Photoelectric
absorption is also a strong function of atomic number, so that, for
example, a silicon detector 1~mm thick will absorb $>$50\% of X-rays
up to 23~keV, while a 1~mm CdTe detector will do the same up to
110~keV.  At these low energies, high spatial resolution is often
desired when there is an imaging system using focusing optics or a
coded mask (see Chapter 12).  This can be accomplished with multiple
small detectors, by pixellating the electrodes of solid-state
detectors, or by using multiple or position-sensitive phototubes to
read out a scintillator (``Anger camera'' configuration, after
inventor Hal Oscar Anger). 

\begin{figure}
\centering
\includegraphics[width=4.5in]{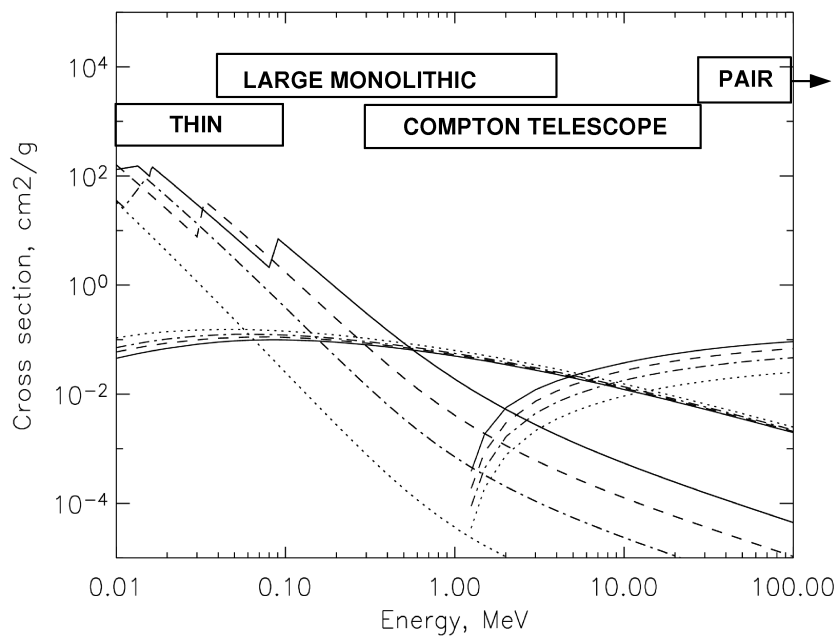}
\caption{Cross-sections for photoelectric absorption (falling), pair
production (rising above 1~MeV), and Compton scattering (flat).  
Data from \citet{Ber98}.  Dotted
line: silicon.  Dash/dot line: germanium.  Dashed line: iodine.  Solid
line: bismuth.  The cross sections of cadmium and tellurium (i.e. 
CdTe and CZT detectors) are similar to iodine.
The approximate energy regimes for basic detector
configurations are shown (thin detectors, large monolithic detectors, Compton 
telescopes, and pair-tracking telescopes).  The pair-tracking regime
extends beyond the plot to hundreds of GeV.}
\label{fig:1}
\end{figure}

At energies of more than about 300~keV, photoelectric cross sections
are small even at high atomic number, and detectors must be made large
enough that photons can Compton scatter in the detector and still be
photoelectrically absorbed afterwards.  Even though the Compton cross
section is nearly independent of atomic number, a high atomic number
is still critical for stopping the dowscattered photon before it
escapes the detector carrying off some of its energy.  A low atomic
number can be desireable for the scattering plane of a Compton
telescope or for a detector or shield designed to stop charged
particles or X-rays and pass $\gamma$-rays through.

In many cases the optimum solution for maximizing sensitivity will be
to have separate detectors for low energies (thin) and high energies
(thick).  Since most cosmic sources have falling energy spectra,
high-energy detectors will generally need larger area than low-energy
detectors in order to reach comparable sensitivity.

Even large monolithic detectors can serve as elements for a coarse
imaging system when placed in a large array.  The INTErnational
Gamma-Ray Astrophysics Laboratory (INTEGRAL) provides two good
examples.  The Spectrometer on INTEGRAL (SPI) \citep{Ve98} has coaxial
germanium detectors with a characteristic size of 7~cm serving as
pixels below a large coded mask (see Chapter 12), and the Imager on
Board the INTEGRAL Spacecraft (IBIS) includes thick fingers of CsI
serving as pixels beneath a finer mask than SPI's \citep{Ub03}.  The
large germanium detectors on the Reuven Ramaty High Energy Solar
Spectroscopic Imager (RHESSI) \citep{Sm02} sit below rotation
modulation collimators (see Chapter 12) that do not require position
sensitivity.

\subsection{Compton and pair tracking telescopes}

At MeV and GeV energies, the physics of the photon interactions in matter
can be exploited to reject background and determine the 
direction of the incoming photon.

In the range of a few hundred keV to tens of MeV,
large-volume detectors with position sensitivity in three dimensions
can record the entire sequence of Compton interactions in the detector
volume.  This allows the direction of the incoming photon to be 
reconstructed and background to be rejected effectively (see Chapter 11).

\begin{figure}
\centering
\includegraphics[width=4.5in]{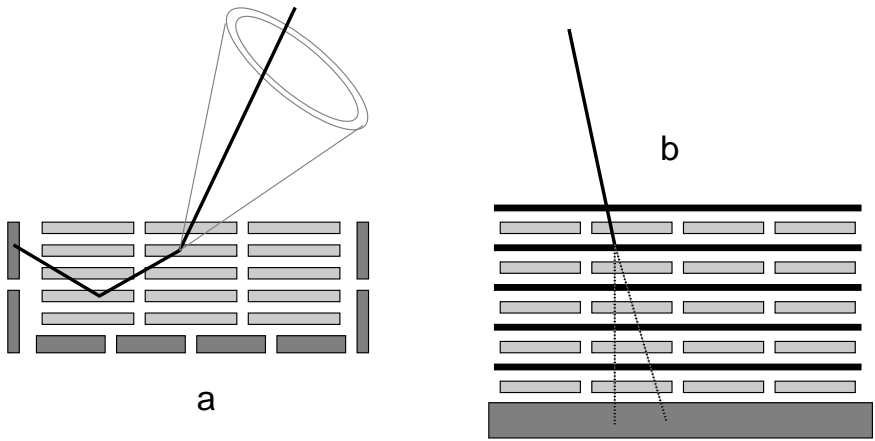}
\caption{
Design concepts for a modern Compton telescope (a) and pair telescope
(b).  The energies and angles between interactions in the Compton
telescope trace the incident photon to an annulus on the sky.  Here the
telescope is shown with low-Z scattering detectors and high-Z absorbing
detectors \citep{Bl02,Zy06,Ta06}, although single-medium instruments are 
also being designed and built \citep{NCT07,Ku04,Ap98}.  See Chapter 11
for more details.
The pair telescope represents the GLAST design \citep{At07}, with 
passive tungsten layers to convert the $\gamma$ into an {\it e+/e-} pair,
silicon detectors to track the pair, and a heavy calorimeter to 
absorb the remaining energy.  Relativistic beaming of the pair and
high spatial resolution allows reconstruction of the
incident photon's direction.
}
\label{fig:2}
\end{figure}

At energies of several tens of MeV and higher, where pair production
is the dominant photon interaction with matter, a pair-conversion
tracking system can be used.  For example, the Large Area Telescope
(LAT) \citep{At07} on the Fermi Gamma-ray Space Telescope
consists of
alternating thin layers of passive tungsten and active silicon strip
detectors (Figure~2b).  Pair production takes place in the tungsten,
since the high atomic number gives a high cross-section (Figure~1).
The electron/positron pair has high enough energy to penetrate
multiple W/Si layers; the penetration depth gives the initial
$\gamma$-ray energy, while the position-sensitive detectors allow the
track to be extrapolated backwards to give the arrival direction of
the $\gamma$-ray.  The highest-energy pairs that penetrate the tracker
are stopped in a calorimeter.  At GeV energies, this extrapolation can
be very precise.  The tracking detectors are not required to measure
deposited energy.

\section{Detector materials}
\label{materials}
\subsection{Scintillation detectors}
\label{scint}

Scintillators can be produced in large, monolithic volumes, and
in a variety of shapes.  They can have low to high atomic numbers (and
therefore stopping power), ranging from plastic scintillators at the
low end to bismuth germanate (Bi$_4$Ge$_3$O$_{12}$ or BGO) at the
high end.  Plastic scintillators can be doped with high-Z atoms, like
lead, to improve their stopping power. 

In inorganic scintillators, ionization produces free electrons that
can move around the crystal until falling back into the valence band.
In {\it activated} crystals, such as NaI(Tl) and CsI(Na), the trace
activator element provides a fast route to the valence band via
intermediate states.  Since the amount of activator is small, the
crystal remains transparent to the scintillation photon emitted when
the activator's excited state decays.  In unactivated crystals such as
BGO, one ion of the pure crystal (Bi$^{3+}$) provides the
scintillation photons, with a large enough shift between its emission
and absorption frequencies that the crystal is still transparent.  In
organic scintillators (plastic and liquid), large molecules are
excited by the passing energetic particles; the scintillation is
produced when they relax to their ground state.  In all cases,
scintillation light can be collected and multiplied by a sensor
such as a photomultiplier
tube (PMT), photodiode, or microchannel plate.  Scintillation light can
be reflected many times before being collected, so the PMT(s) need
not have a direct line of sight to every part of the detector.  In fact,
uniformity of light collection -- and therefore energy resolution --
is sometimes improved by treating the
surfaces of the best-viewed parts of the detector so that they reflect
scintillation light poorly.

In space applications, extremely high energy deposits, up to many GeV,
can occur in a detector due to the passage of a cosmic-ray iron
nucleus (or other heavy element) or the spallation
of a nucleus in the detector by any cosmic ray.  NaI \citep{Fi77} and
CsI \citep{Hu78} scintillators display phosphoresence
in which a fraction of the light is emitted over a much longer time
(hundreds of milliseconds) than the primary fluorescence.  Thus, when
a particularly large energy deposit occurs, the crystal can glow
rather brightly for up to a second; the effect on the data
depends on the design of the electronics.

Pulse shape analysis, whether by analog electronics or via flash
digitization of the PMT signal, has multiple uses for scintillators.
If two different scintillators have very different light decay times,
they can be read out by a single phototube if they are sandwiched
together, with the energy deposited in each still separable in the PMT
signal.  This can be used as a form of active shielding to veto
charged particles or photons that interact in both
scintillators.  This configuration is called a ``phoswich'' (phosphor
sandwich).  Pulse shape analysis can also be used to distinguish
neutron and cosmic-ray ion interactions from the interactions of
photons or electrons in CsI \citep{Bi61} and plastic and
liquid scintillators \citep{Mo92}.

Table~1 summarizes some properties of the scintillators most commonly
used in space, as well as two particularly promising lanthanum halide
scintillators that have recently become available. These new materials
have good stopping power and excellent energy resolution (3--4\% FWHM
at 662~keV versus about 7\% for the industry standard NaI).  They have
a moderate internal radioactivity giving a count rate of $\sim$ 1.8
cm$^{-3}$s$^{-1}$, mostly from $^{138}$La \citep{Ow07}.  This is of
the same order as the background from other sources that an unshielded
detector would receive in low-Earth orbit, but could dominate the
background if the detector is well shielded (see section 4.1 below).

\begin{table}[t]
\caption{Properties of Some Scintillators}
\label{tab1}       % Give a unique label
% For LaTeX tables use
\begin{tabular}{lllllll}
\hline\noalign{\smallskip}
Material & Light Yield & Decay Time & Peak $\lambda$ & Density & Max. & Notes  \\
 & (photons/keV) & (ns) & (nm) & (g/cm$^3$) & Atomic \# & \\
\tableheadseprule\noalign{\smallskip}
NaI(Tl) & 38 & 230 & 415 & 3.67 & 53 & {\it 1} \\
CsI(Na) & 39 & 460, 4180 & 420 & 4.51 & 55 & {\it 2} \\
CsI(Tl) & 65 & 680, 3340 & 540 & 4.51 & 55 & {\it 2} \\
BGO & 8.2 & 300 & 480 & 7.13 & 83 & {\it 3}\\
GSO & 9.0 & 56, 400 & 440 & 6.71 & 64 & {\it 4}\\
BC--408 & 10.6 & 2.1 & 425 & 1.03 & 6 & {\it 5} \\
LaBr$_3$(Ce) & 63 & 16 & 380 & 5.29 & 57 & {\it 6}\\
LaCl$_3$(Ce) & 49 & 28 & 350 & 3.85 & 57 & {\it 7}\\
\noalign{\smallskip}\hline
\end{tabular}

{\it 1} Good energy resolution. Hygroscopic; phosphorescent; 
Susceptible to thermal shock.\\
{\it 2} Slightly hygroscopic; phosphorescent.  Denser, less brittle than NaI.
Pulse-shape discrimination of particle types is possible.\\
{\it 3} Excellent stopping power; inferior energy 
resolution; easily machined; non-hygroscopic. \\
{\it 4} Gd$_2$SiO$_5$; non-hygroscopic.  Used in the Hard X-ray Detector (HXD)
on Suzaku \citep{Ta07}. 
{\it 5} A commonly used plastic. \\
{\it 6} New material; some internal background from
radioactivity. Best energy resolution, good stopping power.\\
{\it 7} Similar to LaBr$_3$, resolution and density not
quite as high.  Large crystals were developed earlier. \\
Data are taken from Tables 8.1 and 8.3 of \citet{Kn00} and from
\citet{Bi07}.
\end{table}

\subsection{Semiconductor detectors}
\label{semi}

In semiconductor detectors, the electrons and holes excited into the
conduction band by the passage of energetic particles are swept toward
opposite electrodes on the detector surface by an applied electric
field.  The image charges induced on one or both electrodes, as they
change with the movement of electrons and holes in the crystal,
provide a small current pulse.  This pulse is generally read by a
charge-sensitive (integrating) preamplifier, followed by a shaping
amplifier.

\begin{figure}
\centering
\includegraphics[width=4.5in]{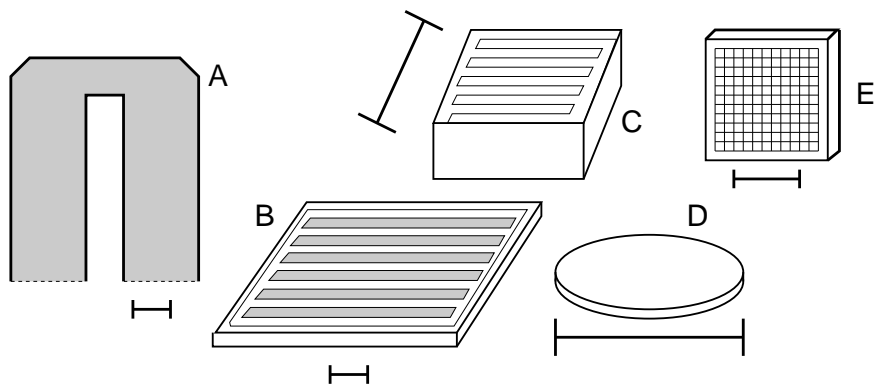}
\caption{
Common electrode configurations for solid state detectors. Each
is shown with a 1~cm marker to indicate the scale of a typical
detector.  A) Coaxial
germanium detector (cross section of a cylindrically symmetrical shape).
Electrodes are on the inner bore and the outer surface; the insulating
surface is at the back (dashed line). B) Silicon strip detector
(typically 300~$\mu$m thick, 1~mm strips).  C) Monolithic CZT
detector with coplanar (interleaved) electrode to null the signal
from the holes and improve energy resolution \citep{Lu94}.  
D) Simple silicon p-i-n
diode with plane electrodes at top and bottom. E) CZT pixel detector.
The pixel and strip detectors are shown with guard rings around the
edge, used to capture any leakage current on the side
surfaces and keep it
away from the electronics chains reading out the volume of the detector.
Pixel and strip detectors have been made from all the common
semiconductor materials (Ge, Si, CZT/CdTe).
}
\label{fig:3}
\end{figure}

Semiconductor detectors are capable of much better energy resolution
than scintillators, since the collection of electrons in the
conduction band is much more complete than the conversion to
scintillation light and light collection in scintillators.  The noise
performance of the electronics must be excellent, however, if the
natural resolution of the detector is to be approached -- i.e., the
limit due to counting statistics of the electron/hole pairs liberated
(including the Fano factor \citep[page 366]{Kn00}).  In the
preamplifier, noise currents are converted to voltage noise
proportional to the detector capacitance \citep[page 33]{Sp05}; thus
detectors with large volume can be expected to show poorer resolution.
Resolution can be preserved at large volumes if the electrode
configuration has intrinsically low capacitance.  Examples of such
configurations are silicon drift detectors \citep{Re85} and germanium
LO-AX$^{\rm TM}$ \citep{Or08} and drift \citep{Lu88} detectors, which
have one large and one small, more pointlike electrode.  Pixellating
the anode and reading out each pixel as a separate detector also
results in low capacitance and excellent noise performance.  Some of
the most common semiconductor detector and electrode configurations
are sketched in Figure~3.

Large energy deposits from cosmic rays in semiconductor detectors do
not carry the risk of long-duration detector response that they do in
scintillators.  It is very important, however, to test the response of
the {\it electronics} to huge energy deposits.   Many designs can become
paralyzed for a significant amount of time by a multi-GeV energy
deposit, or else produce false counts by triggering on ringing.
Interactions this energetic do not occur in the laboratory, where the
only cosmic ray particles are muons, so they ought to be simulated
either at an accelerator or with a pulser before any
electronics design is declared ready for space flight.

An extensive treatment of semiconductor detectors and their electronics
is given by \citet{Sp05}.

\subsubsection{Germanium}
\label{Ge}

Germanium detectors are preeminent for spectroscopy in the range from
hundreds of keV to a few MeV.  Germanium crystals can be
grown in large volumes at extremely high purity, with single detectors
up to 4~kg \citep{Sang99}.  High purity guarantees that both
electrons and holes can move untrapped through the whole crystal
volume, and that the detector volume can be depleted of charge
carriers due to impurities by a manageable applied field of
$\sim$500--1000~V/cm.  The small bandgap gives good counting statistics 
for the liberated electron/hole pairs, but requires low
operating temperatures -- below 130~K, and preferably
much lower -- to prevent thermal excitation 
into the conduction band and a large leakage current 
\citep[pg. 414]{Kn00}.

Energy resolution of $\sim$0.3\% FWHM at 662~keV can be achieved with a
good detector and optimized electronics; this compares to a
corresponding value of $\sim$7\% for NaI(Tl) and $\sim$3--4\% for the new
lanthanum halide scintillators.  Only if this high resolution is
scientifically important should germanium be considered.  But high
resolution should be considered important any time narrow lines are
being observed, even if the exact profile of the line to be observed
is {\it not} needed, and even if it is {\it not} necessary to separate
nearby lines.  Since $\gamma$-ray observations in space are often
dominated by background, a good energy resolution reduces the amount of
background against which the signal of a narrow $\gamma$-ray line is to
be detected, greatly increasing sensitivity.

Operation of germanium detectors in space is challenging due to the
need to keep them cold.  Passive cooling is possible if a large area
and solid angle of radiator can always be presented to deep space
(avoiding the Sun, and, in low-Earth orbit, the Earth as well).  This
technique was used for the Transient Gamma-Ray Spectrometer (TGRS) on
the Wind spacecraft \citep{Se96} and the Gamma-Ray Spectrometer on
Mars Odyssey \citep{Bo04}.  Cryogens can also be used, but they have a
large mass and limited life; the germanium spectrometer on HEAO--3 ran
out of cryogen after a pioneering 154-day mission \citep{Ma80}.  Some
recent instruments have relied on Stirling-cycle mechanical coolers
\citep{Ve98,Sm02,Go07}.  The design of the radiator for waste heat is
still critical in that case, but the requirements are not as severe as
for passive cooling.  RHESSI, for example, often has the Earth nearly
filling the field of view of its radiator for part of its orbit.
Cryocoolers can be very expensive to qualify for space flight and this
should not be underestimated in mission planning.

All the germanium detectors mentioned above were in the closed-end
coaxial configuration (Figure 3a), with an outer contact on the sides
and across the front face of the detector, an inner contact lining a
bore that goes most of the way through the crystal, and an intrinsic
(insulating) surface on the back face.  This configuration is a
compromise between large volume, low capacitance (compared to two flat
electrodes on either side of a comparable crystal) and the lowest
possible distance between the electrodes (to keep the depletion
voltage manageable).  Thick germanium strip detectors are also being
developed for Compton telescopes and other applications that require
position sensitivity, and have already flown on the Nuclear Compton
Telescope balloon payload \citep{NCT07}.  The x and y positions are
measured by localizing the charge collection to individual strips on
each electrode (the strips on one side run perpendicular to those on
the other), while the z position is measured by the relative arrival
times of the electrons and holes at their respective electrodes
\citep{Am00}.

\begin{figure}
\centering
\includegraphics[width=4.5in]{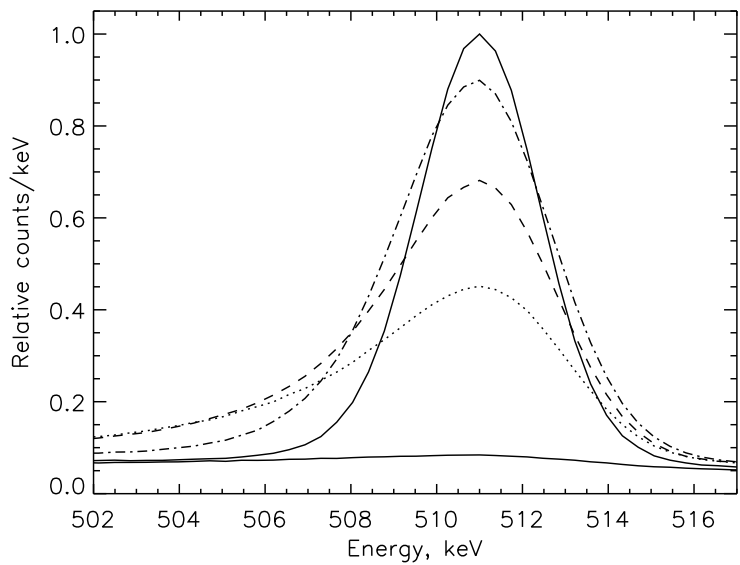}
\caption{Effect of radiation damage on $\gamma$-ray spectroscopic
performance of a coaxial germanium detector.  The 511~keV background
line from positron annihilation is shown in data from the RHESSI
satellite taken from 2002 to 2007.  The symmetrical, narrow peak is
from the start of the mission.  The next two lines (dot-dash and
dashed) show the effect of moderate to severe damage on the line
resolution due to hole trapping.  The last two lines (dotted and the
nearly flat solid line at the bottom) show the loss of effective area
at very severe levels of damage due to volumes in the crystal that are
no longer depleted (active).  At this point, the RHESSI detectors were
annealed.  In general, an anneal would be performed at a much earlier
stage of damage.
}
\label{fig:4}
\end{figure}

In addition to cooling, the other particular challenge for germanium
detectors in space is radiation damage.  Defects in the crystal
lattice caused by nuclear interactions of protons and neutrons create
sites that can trap holes as they drift through the crystal.
Electrons in the conduction band are not comparably affected.  Since
germanium detectors are designed to use both the electron and hole
signals, hole trapping reduces energy resolution.  Because protons
lose energy rapidly by ionization, they must have high energy to
penetrate the layers of passive material around the detector (e.g.
the cryostat) and penetrate beyond the outer surface of the crystal.
Depending on the detector or cryostat configuration, the lower limit
on relevant proton energy is on the order of
100 MeV.  Neutrons, on the other hand, can penetrate the full volume
of the crystal regardless of their energy, and are relevant even at a
few MeV \citep{Le03}.  Protons below 100 MeV can convert to neutrons
via spallation in spacecraft materials and therefore still cause damage.

Strategies to reduce the effect of radiation damage include choice of
orbit, operating procedures, detector geometry, shielding, and
annealing.  

In low-Earth orbit, most radiation damage will come from radiation
belt protons seen during passage through the South Atlantic Anomaly
(SAA).  This can be avoided if the orbit is equatorial (inclination
less than about 10$^{\rm o}$).  This is the most benign orbit
available, since the magnetosphere also protects the instrument from
solar energetic particles and a large fraction of Galactic cosmic
rays.  In high-Earth orbit or interplanetary space, damage from
Galactic cosmic rays usually dominates \citep{Ku99}, unless a large
solar energetic particle event occurs, in which case a large amount of
damage can be inflicted in a short time \citep{Pi07,Ow07}.  An orbit
that spends much of its time in the heart of the radiation belts has
by far the highest dose of all, and would certainly prohibit the use
of germanium.  The Space Environment Information System (SPENVIS)
webpage \citep{He04} is an extremely valuable resource for estimating
the irradiation by radiation-belt and solar protons in
various orbits. 

A good choice of detector geometry can limit the severity of the
effect of radiation damage by limiting the amount of germanium that
the holes must traverse.  In the coaxial configuration, most of the
volume is near the outside of the detector.  Thus, by applying
negative high voltage (HV) to the outer contact, the holes are made to take
the shorter path for the majority of interactions \citep{Pe79}.  The
result is a line shape with a sharp peak and a long tail (due to the
few interactions near the bore).  This is shown in Figure~4.  This
polarity provides good uniformity of field within the crystal for
slightly n-type material.  The opposite polarity (used with p-type
material) will show broadening of the line much earlier and more
severely as the holes migrate all the way to the central bore.

Shielding the detectors can be very effective at blocking solar and
radiation-belt (SAA) protons, but not cosmic rays, which are much more
energetic and penetrating.  Very often, shielding is also desired to
reduce background (see section 4.1 below).

Keeping the detectors very cold reduces the amount of trapping for a
given radiation dose as long as the detector is never warmed up
\citep{Br91}.  Raising the HV on a damaged detector, when
possible, can reduce the effect of damage somewhat \citep{Ku99}.  If
large currents are passed through a damaged crystal, many of the hole
traps will fill, and the effect of radiation damage will be much less
severe for a few minutes, until these traps empty again.  This occurs
when a spacecraft passes through the SAA.  The opposite effect occurs
for damaged n-type germanium detectors when the HV is turned off: when
it is turned back on, there are temporarily {\it more} unfilled traps
than in equilibrium, and the resolution will be degraded until an
equilibrium between detrapping and hole production is reached on the
same timescale of minutes \citep{Ko95}.

Even if all these factors are taken into account in design, virtually
any germanium detector that is {\it not} in a low-Earth, equatorial
orbit will have to be annealed.  When the crystal is heated to well
above operating temperatures, many of the damage sites become
de-activated (not repaired, since the anneal temperatures are far too
low to actually move atoms around in the lattice).  The mechanism is
not well understood.  The literature includes many small-scale
experiments that don't give a good, overall formula for estimating the
efficacy of the anneal process given detector type, damage history,
and anneal temperature and duration.  Temperatures of 
50$^{\rm o}$ to 100 $^{\rm o}$C and durations of days to weeks are 
typical of operations in space
\citep{Lo05,Bo04}; when in doubt, the longer and warmer, the better
\citep{Br91}.  Some annealing does take place at room temperature
\citep{Ra84} but takes longer to be effective, and cannot eliminate
trapping completely.

\subsubsection{Silicon}
\label{Si}

Silicon can be grown in large volumes but not to as high a purity as
germanium, and is therefore harder to deplete.  Silicon detectors are
therefore generally thin (typically 300~$\mu$m), and used for purposes
where that is appropriate.  The thickest silicon detectors (up to
$\sim$1~cm) are made from slightly p-type material and have lithium
ions drifted through the crystal bulk to compensate the intrinsic
impurities, a technique formerly used for germanium before high-purity
material became available.

Small, simple planar p-i-n detectors are often used for X-ray
detection up to a few 10s of keV, as in the top detector layer of the
hard X-ray instrument on the Suzaku spacecraft \citep{Ta07}. Small Si
drift detectors (SDDs), in which the field is shaped to lead the electrons to
a small collecting contact, show improved resolution over p-i-n
detectors due to their smaller capacitance \citep{Re85} and can also
provide position sensitivity when the drift time is measured in the
electronics.

Large, thin Si strip detectors single or double sided) can be used
when position resolution is important but energy resolution and low
energy threshold are not, such as in the silicon tracker on the Fermi
Gamma-ray Space Telescope, where the requirement is to register the
passage of high-energy electrons.  Thick Si strip detectors have been
proposed for a Compton telescope operating in a mode where a final
photoelectric absorption is not necessary \citep{Ku04}.

Si detectors benefit from cooling to reduce leakage current, but at a
more modest level than germanium (temperatures of --20$^{\rm o}$ 
to 0$^{\rm o}$C).  This can be accomplished by a careful passive cooling design
or the use of simple thermoelectric (Peltier) coolers.

\subsubsection{Cadmium Telluride and Cadmium Zinc Telluride}
\label{CZT}

Cadmium telluride (CdTe) and cadmium zinc telluride (Cd$_{\rm
1-x}$Zn$_{\rm x}$Te) offer two advantages relative to germanium: they
can be operated at room temperature and they have better
photoelectric stopping power.  It is difficult to grow large
crystals of high quality, and the largest detectors available
are 1--4~cm$^3$ \citep{Ch08}.  Efficient detection in the MeV range
therefores require a three-dimensional array of detectors 
\citep{Na08} to take
the place of a single large germanium coaxial detector, with very
careful control of passive material within the detector volume to
prevent undetected Compton scatters.

When energy response greater than 30~keV is needed but it is not
necessary to go above a few hundred keV, a single layer of CZT/CdTe
detectors is often the best choice.  If pixels of a few mm or larger
are desired, an array can be made out of individual detectors, as was
done for the Burst Alert Telescope (BAT) \citep{Ba05} on the Swift
mission (CZT) and the front detector layer of the IBIS imager on
INTEGRAL (CdTe) \citep{Le96}.  For smaller pixels, it can be
advantageous to pixellate the electrode on one side of a larger
detector and read signals out of each pixel.  Not only does this
provide greater position resolution with smaller gaps, the small
pixels have very low capacitance and excellent energy resolution.
Pixellated CZT detectors have been used on two hard X-ray focusing
balloon payloads: the High Energy Focusing Telescope (HEFT)
\citep{Bo01} and the INternational Focusing Optics Collaboration for
$\mu$Crab Sensitivity (InFOC$\mu$s) \citep{Tu05}.  For InFOC$\mu$s,
signal traces were routed away from the detector to the ASIC
electronics, while for HEFT the preamplifiers were put onto the ASIC
with the same spacing as the detector pixels and bump-bonded directly
to the detector.  The HEFT detector technology is being adapted
for the Nuclear Spectroscopic Telescope ARray (NuSTAR), an
upcoming NASA Small Explorer satellite using focusing optics
\citep{Ha05}.

CZT and CdTe (and other compound semiconductors) differ from
germanium and silicon in that holes are much less mobile than electrons,
and suffer trapping even in crystals that are not radiation-damaged.
Thus the best energy resolution is obtained when only the electrons
contribute to the energy signal.  This is not possible for a simple
planar configuration (a rectangular detector with plane electrodes on 
opposite sides), but there are a number of ways to improve the 
situation.  Coplanar grid \citep{Lu94,Am97} and 
pseudo-Frisch grid \citep{Mc98} electrode
configurations can cancel most of the contribution of the holes to the 
signal for thick (1 or 2~cm) single detectors, resulting in
an electron-only signal that recovers the excellent energy resolution
of a very thin detector.  If a pixellated
detector is desired for spatial resolution or low capacitance anyway,
pixellating the anode also ensures that the electrons
contribute most of the detected signal as they get very near the
anode (the ``small pixel effect'').  For a thick CZT or Cd/Te detector,
it is also possible to measure the depth of the interaction within 
the crystal by measuring the risetime of the current pulse and use
this information to correct the energy measurement either by analog
or digital techniques downstream \citep{Ri92}.  This technique was
used for INTEGRAL/IBIS \citep{Le96}.

\section{General Considerations}
\label{gen}
\subsection{Background}
\label{bkg}

Since most $\gamma$-ray detectors in space are not detecting focused
radiation, count rates are often dominated by background rather than
signal. Bright transient events such as cosmic $\gamma$-ray bursts,
solar flares, and terrestrial $\gamma$-ray flashes are often
source-dominated, but for most applications both {\it reducing} and
{\it estimating} background levels is important.

For an unshielded detector in low-Earth orbit, the dominant sources of
continuum background are the cosmic diffuse radiation (below about 150
keV) and the ``albedo'' glow of $\gamma$s from the Earth's atmosphere
due to interactions of cosmic rays (above 150 keV) \citep{De03,Ge92}.
From a few hundred keV to a few MeV, radioactivity in detectors
themselves and, to a lesser extent, in passive materials nearby can be
a significant source of both line and continuum background.
Radioactivity can be natural (e.g. from $^{40}$K and the daughters of
$^{238}$U), or induced by cosmic rays, solar or radiation-belt
protons, or neutrons created in the spacecraft or in the Earth's
atmosphere.  Induced radioactivity can be prompt when short-lived
nuclear states have been excited, or have a half-life from seconds to
years.  Above a few MeV, the dominant component is likely to be from
minimum-ionizing cosmic rays clipping the corners of the detector.
Recently, there has been a great deal of effort put into refining
tools for simulating the expected backgrounds \citep{De03,We05} (see
Figure~5).

\begin{figure}
\centering
\includegraphics[width=4.5in]{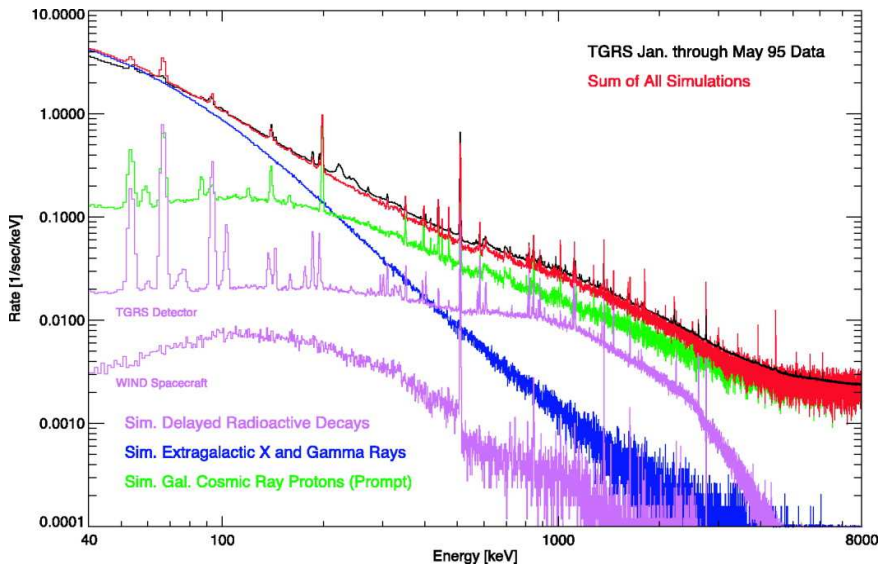}
\caption{
State-of-the-art {\it a priori} modeling of $\gamma$-ray instrumental
background: data from Wind/TGRS and a simulation of
all important background components using MGGPOD.
From \citet{We05}.
}
\label{fig:5}
\end{figure}

{\it Reducing} background should first be approached by selecting the
right detector thickness (no thicker than necessary) and material (one
that is not intrinsically radioactive and does not become badly so
when exposed to cosmic rays on orbit).  But if the instrument is not
meant to observe the entire sky, there should usually also be some
shielding.  Below $\sim$ 100~keV, passive shielding can be adequate.
It is often arranged in a ``graded'' configuration, with a high-Z
material like lead or tungsten on the outside followed by one or two
layers of lower-Z material, each of which is meant to stop K-shell
X-rays from the previous layer before they reach the detector.

At Compton-dominated energies (see Figure~1), thick passive material
would be necessary to stop incoming $\gamma$s.  In space, however,
such a shield can actually create more background than it stops, due
to reprocessing of incident cosmic rays into neutrons, $\gamma$s, and
multiple charged particles.  Thus graded-Z and other passive shields
shouldn't be more than a few millimeters thick.  However, {\it active}
shielding with several centimeters of inorganic scintillator can be
very effective.  In this case, cosmic rays are vetoed along with their
daughter particles produced in the shield, and only a single
interaction is necessary to veto a background photon, even if it then
interacts in the central detector. An active shield will also veto
photons from the target that interact in the central detector but
scatter out of it (``Compton shield'' mode).  Even active shields
produce background via neutron production \citep{Na96}.  A study for
INTEGRAL/SPI found that 5~cm of BGO was the optimum shield thickness
for its orbit \citep{Di96}.

At the highest energies, a thin, active plastic shield can veto the
prompt components due to cosmic rays: clipping of the detector by the
cosmic rays themselves and prompt nuclear de-excitations in passive
materials near the detectors. But it should be established that these
background components will be important in the energy range of
interest before choosing to use a plastic veto.  The ability to veto
charged particles that are {\it not} cosmic rays is desirable for an
orbit outside the Earth's magnetosphere (for solar particles) or a
low-Earth orbit that goes to high magnetic latitudes (for
precipitating outer-belt electrons and, for nearly polar orbits, solar
particles as well).

{\it Estimating} background is less important for detectors in imaging
configurations (e.g. coded mask, rotating grid, or Compton telescope)
that have ways to reject background based on incident direction. In
these cases, it is enough to predict the background accurately enough
to have confidence in the instrument's sensitivity.  But for
non-imaging detectors, it may be necessary to know the background to 
1\% or better to study faint sources.  This cannot be done
via {\it a priori} modeling, if only because cosmic ray fluxes
fluctuate much more than this.  Instead, background is subtracted by
finding a period of time when the source is not visible but the
background is expected to match that during the observation.  For
highly collimated instruments, this is best accomplished by
``chopping'' between the source position and an empty field nearby, as
was done with the Oriented Scintillation Spectrometer Experiment
(OSSE) on CGRO and the High-Energy X-ray Timing Experiment (HEXTE) on
RXTE \citep{Jo93,Ro98}.  For uncollimated or wide-field instruments
viewing a transient event like a cosmic $\gamma$-ray burst or a solar
flare, time intervals just before and after the event, or (in
low-Earth orbit) $\pm$ 15 orbits ($\sim$ one day) away often provide
an excellent background measurement.  The case of a non-chopping
instrument measuring a non-transient source is the most difficult.  A
variety of techniques can be mixed, combining observations and
modeling, including the use of the Earth as an occulter and the
generation of background databases incorporating large amounts of data
from throughout the mission.  Such a database can be used to extract
the dependence of background on parameters such as orbital position
and cosmic ray flux \citep{Ja06,Sm96}.

\subsection{Livetime}
\label{live}

Instrumental livetime is generally of concern only when background is
not -- i.e. when very bright cosmic, solar or terrestrial transients
are of primary interest.  In these cases, all stages of the signal
chain should be analyzed to make sure that the highest expected count
rate can be recorded.  The intrinsic response time of the detector
material (scintillation light decay or electron and hole drift times
in a solid state detector) may be a consideration in detector choice,
but only if pulse shaping times and throughput in the rest of the
electronics can be designed to keep up with the detector's capability.
Large detectors can be pixellated or replaced with many small ones to
reduce deadtime, at the expense of an increase in the number of
electronics chains needed.  The deadtime caused by an active shield
veto should always be estimated, particularly when the instrument is
very large or will be studying bright transients.  When scattering
between detector elements is part of the source detection (such as in
a Compton telescope or $\gamma$ polarimeter), the frequency with which
two independent background or source interactions will fall within the
instrument's coincidence time window by chance and be mistaken for a
scatter should always be calculated.

\begin{figure}
\centering
\includegraphics[width=4.5in]{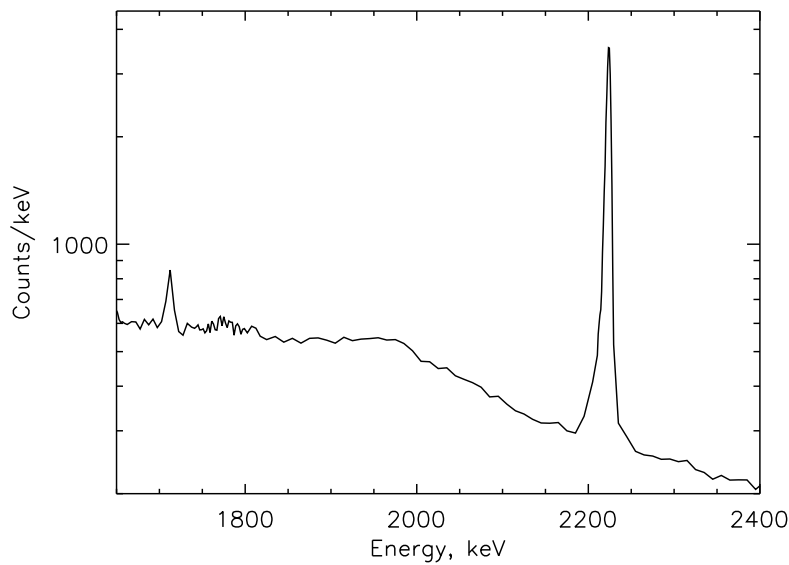}
\caption{
RHESSI spectrum of the solar flare of 28 October, 2003.  The emission
in this energy range is dominated by the response to the 2.223~MeV
line, which includes the photopeak, the Compton continuum from photons
that scatter nearly 180$^{\rm o}$ out of the detector (cutting off
around 2000 keV), and the first 511~keV escape peak (1712 keV).  There
is an underlying, falling continuum due to other flare components as
well.
}
\label{fig:6}
\end{figure}

\subsection{Spectral Response}
\label{resp}

Lastly, it is important to understand the energy response of any
detector design due to the physics of the high-energy photon
interactions.  Incomplete collection of the incident photon energy is
important for both line and continuum spectroscopy, but is most
obvious to the eye when a narrow line is being observed.  At energies
where Compton scattering becomes important, a Compton continuum below
the incident energy appears in the spectrum due to scattering either
into or out of the detector.  At low energies (within a factor of
$\sim$2 the K-edge of the detector material), a K-shell X-ray escape
peak appears since absorption occurs very close to the surface. At MeV
energies, two escape peaks appear corresponding to the escape of one
or both 511~keV photons following pair production and annihilation of
the positron.  Figure~6 shows the photopeak line, Compton backscatter
feature, and first 511~keV escape peak from a RHESSI observation of
the 2.223~MeV line from a solar flare, from neutron capture by a
proton producing deuterium.

These effects -- combined with the blurring effect of finite energy
resolution -- combine to make up the ``response matrix'', the function
that maps the input spectrum of incoming photons to the output
spectrum of detector counts.  Good stopping power (high atomic number)
and an active Compton shield can keep the diagonal components of this
matrix dominant, making interpretation of the spectrum easier.  It is
not possible to unambiguously invert a nondiagonal instrument response
matrix and calculate a unique photon spectrum given the observed count
spectrum.  The usual practice is to convolve models of the expected
spectrum with the response matrix and compare the results to the
observed count spectrum.

Monte Carlo tools such as GEANT should be used to model and predict
the response matrix for any new design.

\section{Outlook}

Germanium (or, at hard X-ray energies, the other solid state
detectors) provides high enough energy resolution for most
astrophysical purposes $>$ 10 keV.  BGO provides stopping power and
large volume up to the mass limit that a launch vehicle can reasonably
haul.  So the current challenges in detector development are combining
high resolution with high volume, and adding fine, 2- or 3-dimensional
spatial resolution for Compton telescopes, coded masks, or imagers.
Germanium strip detectors provide an appealing compromise of very high
spatial and energy resolution with moderate detector volume and
stopping power.  The key development issues are the difficulty, power
and expense of cooling and minimizing the amount of passive material
in and around the array, material needed for thermal control as well
as mechanical and electrical connections.  When very high spatial and
spectral resolution are not required, the new lanthanum halide
scintillators are becoming a popular option in proposals for space
instruments, have spectral resolution better than other scintillators
and very good stopping power.  They are appealing if their intrinsic
background can be tolerated.  Recently, high-performance Anger camera
prototypes have been developed using SDDs to read out a large
scintillator \citep{Le08, La08}; SDDs have also been coupled to a
physically pixellated scintillator \citep{Ya07}.  These technologies
may provide an interesting alternative to pixellated semiconductors
for moderate to high spatial resolution.

Development is also always in progress on other technologies
that are promising but still present technical challenges, 
such as liquid xenon \citep{Ap98, Ap08}, semiconductors
such as TlBr \citep{Ch08, Hi08} and HgI$_2$ \citep{va07}, 
and scintillators such as SrI$_2$ \citep{Wi08, Ha08} and 
BaI$_2$ \citep{Ch07}.

%\begin{acknowledgements}
I would like to thank Mark McConnell for very helpful suggestions on
the manuscript of this chapter.
%\end{acknowledgements}

\end{sloppypar}

\end{document}